\documentclass[a4paper,12pt]{article}
\usepackage[cp1251]{inputenc}
\usepackage[english]{babel}
\usepackage[dvips]{graphicx}
\graphicspath{{.}}

\begin{document}
\mathsurround=2pt \sloppy
\title{\bf Pairbreaking effect of correlated impurities in a superfluid Fermi liquid}
\author{I. A. Fomin  \\
{\it P. L. Kapitza Institute for Physical Problems}\\ {\it Russian Academy of Science},\\{\it
Kosygina 2,
 119334 Moscow, Russia}}
\maketitle
\begin{abstract}
The conventional theory of superconducting alloys does not take into account a discreet character of impurities. Experimental data for superfluid $^3$He in aerogel and for some of high-$T_c$ superconductors reveal a significant discrepancy between the observed temperatures $T_c$  of their transitions in the superfluid or superconducting state and that predicted by the theory.
Here a theoretical scheme is presented for finding corrections to the  $T_c$ originating from spatial correlations between impurities. Analysis is limited to the Ginzburg and Landau temperature region. 
The shift of $T_c$ with respect to the pure material is represented as a series in concentration of the impurities $x$. In the first order on $x$ the conventional mean-field result for the lowering of $T_c$ is recovered. Contribution of correlations enters the second order term. It is expressed via the structure factor of the ensemble of impurities. For superfluid $^3$He in silica aerogel the sign of the correction corresponds to an enhancement of the  $T_c$ so that the resulting pairbreaking effect of impurities is weakened. When correlation radius of impurities $R$ exceeds the coherence length of the superfluid $\xi_0$ the contribution of correlations to the shift of $T_c$ acquires a factor $\sim(R/\xi_0)^2$ and the weakening of the pairbreaking effect becomes  appreciable. The presented scheme is applied to the superfluid $^3$He in aerogel.
\end{abstract}
\section{Introduction}
Impurities can significantly change properties of a superconductor. E.g. it is possible to change a type of superconductor from first to second by an increase of concentration of impurities. In case of unconventional Cooper pairing impurities lower transition temperature $T_c$.  The word "unconventional" means that at this type of Cooper pairing not only gauge symmetry but some other symmetries of a normal phase are broken. Transition temperature is one of the most important thermodynamic characteristics of a superconductor. It is also one of the first quantity to be measured in experiments. Abrikosov and Gorkov (AG) \cite{AG} theory of superconducting alloys renders quantitative description of the lowering of the transition temperature and of other properties of superconducting alloys in terms of one parameter - a transport mean free path $l_{tr}$ of electrons in a given material, which in its turn is inversely proportional to the number of impurities per unit volume $n$, or their concentration $x$.  In the AG-theory it is assumed, that distribution of impurities is completely random, i.e. their positions are not correlated. Effect of correlations on thermodynamic properties of superconductors is of the second order on concentration and in most cases it can be neglected in comparison with the principal mean field effect, which is linear in $x$, but there may be special reasons for an enhancement of contribution of correlations.

In the last two-three decades many unconventional superconductors were discovered, including high-T$_c$ materials.
Doping with impurities is significant part of a processes of their preparation. The pairbreaking effect of impurities in high-$T_c$ superconductors is observed experimentally \cite{attfield,fujita}, but the lowering of $T_c$ in many cases is smaller than that, predicted by the AG theory (cf. \cite{franz} and references therein). This tendency is interpreted in Ref. \cite{franz} as a manifestation of inhomogeneity of the condensate which can be appreciable in superconductors with a very short coherence length.

Doping of metallic superconductors changes not only concentration and distribution of scattering centers but also other parameters like concentration of charge carriers, lattice constant etc., which in their turn change the $T_c$. A special effort has to be paid to separate effects of different factors.
The mentioned difficulties are not present in another physical object, where effect of correlation of impurities is observed -- it  is superfluid $^3$He in aerogel. Liquid $^3$He becomes superfluid at temperatures of the order of 1 mK. $^3$He is unconventional superfluid -- Cooper pairs here are formed in a state with angular momentum $l=$1 and spin $s=$1, so that the rotational symmetries both in spin and in the orbital space are broken. Properties of pure (or bulk) superfluid $^3$He are thoroughly investigated experimentally and most of them are quantitatively interpreted \cite{VW} so that changes introduced by impurities can be accurately separated. All that makes liquid $^3$He a favorable object for a study of effect of impurities, except that introduction of impurities in the superfluid $^3$He is not a trivial task. Possible floating impurities at these low temperatures stick to the walls of a container. To remain in the bulk of the liquid impurities have to form a self supporting structure. As this in experiments high porosity aerogels are used \cite{parp1,halp1,askh}. Majority of data is obtained with silica aerogel. It consists of very thin (diameter $d\approx$ 3 nm) strands of SiO$_2$. Scattering of quasi-particles by the silica strands renders a finite mean free path $l_{tr}$ and lowers a temperature of transition in the superfluid state. Instead of concentration aerogels are traditionally characterized by the \emph{porosity} $P$, i.e. by the fraction of space which is not occupied by the strands. In most of the experiments with $^3$He silica aerogels with $P\approx$98\% are used. In theoretical models the strands forming aerogel are often treted as chains of silica balls of a radius $\rho\approx d/2$, which play the part of scattering centers. The centers can be treated as impurities in the conventional theory as long as $\rho\ll\xi_0$ where $\xi_0$ is the coherence length of the superfluid.

Essentially new property of aerogel is that for forming a rigid structure positions of the centers have to be correlated in space. The correlation function is an intrinsic property of aerogel, it does not depend on a substance filling its pores, and it can be directly measured by $X$-ray scattering.  Presence and importance of correlations was realized at the very beginning of investigation of $^3$He in aerogel \cite{parp2}. In particular correlations were suggested to be responsible for deviations of the observed dependence of $\frac{\delta T_c}{T_c}$ on $\frac{\xi_0}{l}$ from that predicted by the AG theory. For curing the disagreement more complicated theoretical models of $^3$He in aerogel were suggested \cite{thuneb}. One of them combines description in terms of impurities with that in term of pores (isotropic inhomogeneous scattering model IISM). This model improves agreement with the data for $^3$He but it has certain drawbacks. The model exploits  uncontrolled approximations and exact physical meaning of fitting parameters is not clear. Another successful model (phenomenological IISM) is an heuristic interpolation between the limiting situations of impurities and pores.   It is not clear also how the IISM can be generalized for application to correlated impurities in superconductors.

In a present paper correlations are treated within the perturbative approach, developed by Larkin and Ovchinnikov  for superconductors with a scalar order parameter \cite{LO}. Effect of quenched disorder is introduced in the Ginzburg and Landau equations via position dependent coefficients. Random deviations of the coefficients from their average values are treated as 
perturbations.  This approach is more standard and more universal than the previously used. A preliminary account of application of  such approach to the superfluid $^3$He in silica aerogel was published before \cite{fomin}. In comparison with Ref.\cite{fomin} here a more general and more realistic formulation is used. In particular the anisotropy of coherence length of $^3$He and possible  anisotropy of scattering cross-sections of impurities are taken into account. Better estimation of a region of applicability of the present approach is given and the process of summation of the perturbation series is fully reconsidered.

\section{Phenomenology}
Pair-breaking impurities make the condensate of  Cooper pairs spatially
nonuniform. In a vicinity of the transition temperature $T_c$ free energy of such superconductor
can be written as a Ginzburg and Landau functional with the
coefficients, which are random functions of coordinate. For a condensate with the
scalar order parameter $\Psi(\textbf{r})$:
$$
\emph{F}_s\{\Psi(\textbf{r})\}=\emph{F}_n+
\int\{a(\textbf{r})|\Psi(\textbf{r})|^2+
\frac{1}{2}b(\textbf{r})|\Psi(\textbf{r})|^4+
c(\textbf{r})|\nabla\Psi(\textbf{r})|^2\}d^3r.           \eqno(1)
$$
In view of further applications
 magnetic field is not taken into account in this functional. For a situation when fluctuations of coefficients
 $a(\textbf{r}), b(\textbf{r}), c(\textbf{r})$ at their average values are small
 Larkin and Ovchinnikov \cite{LO}  have analyzed the
 effect of inhomogeneities on thermodynamic and electro-magnetic properties of a superconductor
with the scalar order parameter treating the fluctuations as a perturbation. According to their analysis
 the most strong effect on the average value of the order parameter and on the shift of the transition
temperature have fluctuations of the coefficient $a(\textbf{r})$, which can be expressed in terms of fluctuations
of the local transition temperature $T_c(\textbf{r})$: $a(\textbf{r})=\alpha(T-T_c(\textbf{r}))$.
Contribution of these fluctuations is singular at $T\rightarrow T_c$. Contributions
of fluctuations of $b(\textbf{r})$ and $c(\textbf{r})$ are regular in this limit, they can be neglected so that the average over ensemble of impurities,  values $\langle b\rangle$ and $\langle c\rangle$ can be used in the equations. If $T_{c0}$ is the temperature of transition in the absence  impurities, then $a(\textbf{r})=\alpha T[\eta(\textbf{r})-\tau]$, where  $\tau=(T_{c0}-T)/T$ and
$\eta(\textbf{r})=(T_{c0}-T_c(\textbf{r}))/T$ is a random local shift of $T_c$. After introducing dimensionless variables
$\Psi=\Psi_0\psi$ with $\Psi_0^2=\alpha T\langle b\rangle$, and  $\xi_s^2=\langle c\rangle/\alpha T$, or  $\xi_s^2=\frac{7\varsigma(3)}{20}\xi^2_0$ the equation for the extremum of the functional Eq. (1) takes a form
 $$
 [\tau-\eta(\textbf{r})]\psi+\xi_s^2\Delta\psi-\psi|\psi|^2=0.                            \eqno(2)
 $$
Global $T_c$ is defined as the highest value of $T$ for which   $\langle\psi\rangle\neq 0$. The signs in the definitions of $\tau$ and $\eta(\textbf{r})$ are chosen so that the linear part of Eq (1) is analogous to the Schrodinger equation of a particle moving in a random potential $\eta(\textbf{r})$ and $\tau$ has a meaning of energy. Superconducting  states correspond to positive $\tau$. For small $\eta(\textbf{r})$ a shift of the transition temperature $\tau_c=(T_{c0}-T_c)/T_c$ can be represented as a perturbation series
 $\tau_c=\tau^{(1)}_c+\tau^{(2)}_c+...$, where $\tau^{(1)}_c=\langle\eta(\textbf{r})\rangle$ and $\tau^{(2)}_c$ depends on the correlation function $\langle\eta(\textbf{r})\eta(\textbf{r'})\rangle$.

The perturbative approach of Larkin and Ovchinnikov can be applied to condensates with a multi-component order parameter as well.
 That requires modification of the free energy functional according to the form of the order parameter. In superfluid $^3$He the order parameter is a complex 3$\times$3 matrix $A_{\mu j}$ \cite{VW}. Its first (Greek) index corresponds to three possible projections of spin and the second (Latin) -- to three projections of the orbital momentum of a Cooper pair. Non-magnetic impurities interact with the orbital part of $A_{\mu j}$ and the additional term in the density of free energy is $f_{\eta}= \eta_{jl}({\bf r})A_{\mu j}A_{\mu l}^*$, where $\eta_{jl}({\bf r})$ is a random tensor field. It is assumed that the impurities preserve $t\to -t$ symmetry, then $\eta_{jl}({\bf r})$ is real and symmetric.

 Like in the scalar case, the perturbation can be separated in the ensemble averaged part
 $\langle\eta_{jl}\rangle=\eta^{(0)}\delta_{jl}+\kappa_{jl}$
 and the fluctuation $\bar{\eta}_{jl}(\textbf{r})=\eta_{jl}(\textbf{r})-\langle\eta_{jl}\rangle$  The isotropic part $\eta^{(0)}$
can be included in the definition of $\langle T_c\rangle$ so that $\tau=\frac{\langle T_c\rangle-T}{T}$ and
$\kappa_{jl}$ is a global anisotropy. The finite anisotropy splits one $T_c$  generally into three $T_c$ corresponding to different components of $l$. As a result in a neighborhood of $T_c$ there may be several transitions into phases with different order parameters
\cite{AI,fom4}. To avoid irrelevant complications in what follows it will be assumed that the ensemble of impurities is isotropic i.e. $\kappa_{jl}=0$. The physical meaning of  $\bar{\eta}_{jl}(\textbf{r})$ is clear from the definition -- its isotropic part
 describes local fluctuations of $T_c(\textbf{r})$, while the anisotropic part $\eta^{(a)}_{jl}({\bf r})\equiv\eta_{jl}({\bf r})-
\frac{1}{3}\eta_{ll}({\bf r})\delta_{jl}$ describes the local splitting of $T_c$ for different components of $l$. With these notations
$$
 F_{GL}=
 $$
 $$N(0)\int d^3r\left\{[
\eta_{jl}({\bf r})-\tau\delta_{jl}]A_{\mu j}A_{\mu l}^*+
\xi_s^2\left(\frac{\partial A_{\mu l}}{\partial x_j}
\frac{\partial A^*_{\mu l}}{\partial x_j}+
2\frac{\partial A_{\mu l}}{\partial x_j}\frac{\partial A^*_{\mu j}}{\partial
x_l}\right)+\frac{1}{2}\sum_{s=1}^5 \beta_sI_s \right\} ,
$$
$$
                                                                                    \eqno(3)
$$
where $I_s$ - are the
fourth order invariants in the expansion of the free energy over $A_{\mu j}$ \cite{VW} and  $\beta_s$, $s=1,...5$ are
phenomenological coefficients.

  For finding of the  $T_c$ in a presence of perturbation  $\eta_{jl}(\textbf{r})$ it is sufficient to keep in the Ginzburg and Landau equation corresponding to the functional (3) only linear with respect to
 $A_{\mu j}$ terms
$$
[\tau\delta_{jl}-\eta_{jl}(\mathbf{r})]A_{\mu l}+\xi^2_s\left(\frac{\partial^2 A_{\mu j}}{\partial x_l^2}+
2\frac{\partial^2 A_{\mu l}}{\partial x_l \partial x_j}\right)=0                           \eqno(4)
$$
and to use a standard perturbation procedure. Fourier transform of
the Green function $G_{jl}(\mathbf{k},\mathbf{k}')$ of Eq. (4) is
developed in the series, graphically represented on Fig. 1. Each
line corresponds to the Green function of the unperturbed Eq. (4)
$$
G^{(0)}_{jl}(\tau,\mathbf{k})=\frac{\hat{k}_j\hat{k}_l}{\tau-3\xi_s^2k^2}+\frac{\delta_{jl}-\hat{k}_j\hat{k}_l}{\tau-\xi_s^2k^2}.        \eqno(5)
$$
and each cross -- to the Fourier transform of the perturbation
$\eta_{jl}(\mathbf{k}_{s+1}-\mathbf{k}_s)$. The series has to be
averaged term by term  over the ensemble of $\eta_{jl}$ and the
result is re-summed into the averaged Green function, which is
spatially uniform:
$$
\langle G_{jl}(\tau,\mathbf{k},\mathbf{k}')\rangle=(2\pi)^3\delta(\mathbf{k}-\mathbf{k}')G_{jl}(\tau,\mathbf{k}).        \eqno(6)
$$

\begin{figure}[t]
\begin{center}
\includegraphics[clip, scale=0.7]{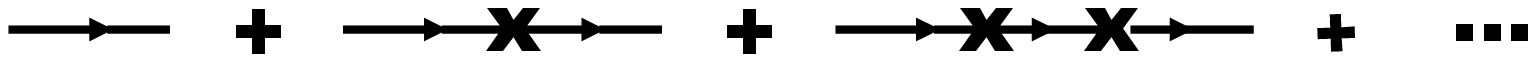}
\end{center}
\caption{Perturbation series for the Green function of Eq. (4)}
\end{figure}

In its turn

$$
G_{jl}(\tau,\mathbf{k})=G^{(0)}_{jl}(\tau,\mathbf{k})+G^{(0)}_{jm}(\tau,\mathbf{k})\langle W_{mn}(\mathbf{k},\tau)\rangle G^{(0)}_{ml}(\tau,\mathbf{k}),       \eqno(7)
$$

where $\langle W_{mn}(\tau,\mathbf{k})\rangle$ is the averaged sum of the series graphically represented on Fig 1 if the thin lines on both ends of each graph are omitted:
$$
\langle W_{mn}(\tau,\mathbf{k})\rangle=\sum_p w^{(p)}_{mn}(\tau,\mathbf{k})      \eqno(8)
$$
If  $\langle W_{jl}(\tau,\mathbf{k})
\rangle$ is small (the criterion will be formulated later), in a principal order
$$
G_{jl}^{-1}(\tau,\mathbf{k})=(\tau-3\xi_s^2k^2)\hat{k}_j\hat{k}_l+(\tau-\xi_s^2k^2)(\delta_{jl}-\hat{k}_j\hat{k}_l)- \left\langle W_{jl}(\tau,\mathbf{k})\right\rangle.             \eqno(9)
$$
Transition temperature $\tau_c$ in a presence of perturbation is found as a pole of $G_{jl}(\tau,0)$
$$
\tau_c\delta_{jl}=\langle W_{jl}(\tau_c,0)\rangle.                               \eqno(10)
$$
In what follows all functions of $\tau$ will be used at $\tau=\tau_c$ and the argument $\tau$ will be suppressed.
In the first order on the perturbation the mean field result is recovered
$$
w^{(1)}_{jl}(0)=\langle\eta_{jl}(\textbf{k}=0)\rangle      \eqno(11)
$$
and in the second order we arrive at a straightforward generalization of the corresponding result of Ref. \cite{LO}:
$$
w^{(2)}_{jl}(0)=-\int\frac{3\delta_{nm}-2\hat{k}_m\hat{k}_n}{3\xi_s^2k^2}
\langle\eta_{jm}(\textbf{k})\eta_{nl}(\textbf{-k})\rangle
\frac{d^3k}{(2\pi)^3}.    \eqno(12)
$$
The variable  $\tau$ does not enter this expression because of the condition (10).
\section{Impurities}
For application of the obtained expressions a form of $\eta_{jl}(\mathbf{r})$ has to be specified. In what follows we assume that perturbation is produced by ensemble of discreet identical impurities situated at random positions $\mathbf{r}_a$. If linear dimensions of each impurity $\rho$ meet the condition $\rho\ll\xi_0$ they can be considered as "small objects" in a sense of the theory of Rainer and Vuorio \cite{Rainer}. According to this theory every impurity acts on the condensate as a localized perturbation with a characteristic size of the order of $\xi_0$.    For example, if the impurity is a ball of the radius $\rho$, which diffusely scatters quasi-particles, the corresponding perturbation at distance $r\gg\varrho$ from the center is given by
$$
\eta_{jl}^{(1)}({\bf
r})=-\frac{\rho^2}{r^2}\hat\nu_j\hat\nu_l\ln\left[
\tanh\left(\frac{r}{2\xi_0}\right)\right],  \eqno(13)
$$
where $\hat\nu_j$ is a unit vector parallel to $\mathbf{r}$.
If $\xi_0\ll l$ the perturbation produced by all impurities can be approximated by the sum of contributions of individual
impurities:
$$
\eta_{jl}(\mathbf{r})=\sum_s\eta_{jl}^{(s)}({\bf r-r_s}).  \eqno(14)
$$
The Fourier transform of $\eta_{jl}(\mathbf{r})$ is
$$
\eta_{jl}(\mathbf{k})=\sum_s\eta_{jl}^{(s)}(\mathbf{k})\exp(-i\mathbf{k}\mathbf{r}_s).  \eqno(15)
$$

For substitution in Eq.(11) we need only $\eta_{jl}(0)=\sum_s\eta_{jl}^{(s)}(0)$. For identical impurities tensors $\eta_{jl}^{(s)}(0)$ differ only by their orientation. In case of uniaxial impurities \cite{Rainer}
$$
\eta_{jl}^{(s)}(0)=\frac{\pi^2}{4}\xi_0\left[\sigma_{tr}^{(i)}\delta_{jl}+\sigma_{tr}^{(a)}(3\hat{a}^{(s)}_j\hat{a}^{(s)}_l-\delta_{jl})\right],     \eqno(16)
$$
where $\hat{a}^{(s)}_j$ is a unit vector in the direction of symmetry axis of the impurity at a point  $\mathbf{r}_s$ and
the transport cross-sections $\sigma_{tr}^{(i)}$ and $\sigma_{tr}^{(a)}$ are expressed through the differential cross-section  of scattering of quasi-particles by the impurity as:
$$
\sigma_{tr}^{(i)}\delta_{jl}+\sigma_{tr}^{(a)}(3\hat{a}_j\hat{a}_l-\delta_{jl})=3\int\frac{d^2\vartheta}{4\pi}\int d^2\vartheta'[\hat{\nu_j}\hat{\nu_l}-\hat{\nu'_j}\hat{\nu_l}]\frac{d\sigma}{d\Omega}(\mathbf{\nu},\mathbf{\nu'}).    \eqno(17)
$$
At the averaging over directions of $\hat{a}^{(s)}_j$ the term, proportional to  $\sigma_{tr}^{(a)}$ in Eq. (14) vanishes. Combining Eqns. (8),(9),(11) and (16) we arrive at the first order correction to the transition temperature
$$
\tau_{jl}^{(1)}=n\frac{\pi^2}{4}\xi_0\sigma_{tr}^{(i)}\delta_{jl}                               \eqno(18)
$$
This expression coincides with the result of AG theory in the first order on the ratio $\xi_0/l_{tr}$. The correction is negative, i.e. the transition temperature decreases.

To find $\tau_{jl}^{(2)}$ we have to substitute in  the r.h.s. of Eq. (12) the explicit form of $\eta_{jl}(\mathbf{k})$. It will be shown that the principal contribution to the integral comes from the region of small $\mathbf{k}$. Then the impurities can be considered as point-like:
$$
\eta_{jl}({\bf r})=\sum_s\eta_{jl}^{(s)}(0)\delta({\bf r-r_s})  \eqno(19)
$$
with the Fourier transform:
$$
\eta_{jl}(\mathbf{k})=\sum_s\eta_{jl}^{(s)}(0)\exp(-i\mathbf{k}\mathbf{r}_s).                        \eqno(20)
$$
Eventually
$$
w^{(2)}_{jl}(\tau,0)=-\left(\frac{\pi^2\xi_0}{4}\right)^2\int\frac{d^3k}{(2\pi)^3}\frac{3\delta_{nm}-
2\hat{k}_m\hat{k}_n}{3\xi_s^2k^2}
$$
$$
\left[\delta_{jm}\delta_{nl}(\sigma_{tr}^{(i)})^2
\langle\sum_{t,s}e^{i\mathbf{k}(\mathbf{r}_t-\mathbf{r}_s)}\rangle+
(\sigma_{tr}^{(a)})^2
\langle\sum_{t,s}(3\hat{a}^{(s)}_j\hat{a}^{(s)}_m-\delta_{jm})(3\hat{a}^{(t)}_n\hat{a}^{(t)}_l-\delta_{nl})
e^{i\mathbf{k}(\mathbf{r}_t-\mathbf{r}_s)}\rangle\right].   \eqno(21)
$$
If impurities are spherically symmetric $\sigma_{tr}^{(a)}=0$ $\eta_{jl}^{(s)}(0)=\eta_0\delta_{jl}$. Distribution of impurities is characterized by one function -- the structure factor
$$
S(\mathbf{k})=
\langle\sum_t e^{i{\mathbf k}(\mathbf{r}_t-\mathbf{r}_s)}\rangle.      \eqno(22)
$$
In the structure factor the contribution of correlations can be separated. The term with $t=s$ is always present, it is equal to unity and when substituted in Eq. (19) renders second order correction to the shift of the transition temperature by non-correlated impurities.  The remaining sum can be written as an integral
$$
S(\mathbf{k})-1=n\int C({\bf r}_t|{\bf r}_s)e^{-i\mathbf{k}(\mathbf{r}_t-\mathbf{r}_s)}d^3r_t,  \eqno(23)
$$
where
 $C({\bf r}_t|{\bf r}_s)$   is  the probability to find a particle in the point ${\bf r}_t$ if there is a particle in the point ${\bf r}_s$. For isotropic distribution of impurities it depends only on a distance $r=|\mathbf{r}_t-\mathbf{r}_s|$.  At $r\rightarrow \infty$ correlations vanish and $C(r)$ tends to a constant. Normalization of $C(r)$ is chosen so that this constant is unity. The unity contributes to $S(\mathbf{k})$ a term, proportional to $n\delta(\mathbf{k})$, which is already taken into account in $\tau_{jl}^{(1)}$. A measure of correlations is $v(r)=w(r)-1$. The structure factor can now be represented as $S(\mathbf{k})=1+(2\pi)^3n\delta(\mathbf{k})+\bar{S}(\mathbf{k})$, where only the last term depends on correlations
$$
\bar{S}(\mathbf{k})=n\int v(r)e^{-i\mathbf{k}\mathbf{r}}d^3r. \eqno(24)
$$
Its contribution to $\tau^{(2)}$ is:
$$
\bar{\tau}_{jl}^{(2)}=-\frac{5}{2\zeta(3)}\left(\frac{\pi}{6}\right)^2 n(\sigma_{tr}^{(i)})^2\delta_{jl}\int_0^{\infty} \bar{S}(k)dk.               \eqno(25)
$$
The X-ray scattering data for silica aerogels \cite{frel,parp2,halp2} show that there exist an interval of $k$: $(1/R)<k<(1/\bar{\rho})$ where  $S(k)\sim (1/k^{D_f})$, i.e. these aerogels have a fractal structure with the fractal dimension $D_f\approx1.7\div 1.9$ depending on a sample. The inverse lower boundary of the fractal interval has to be identified with the correlation radius $R$, the inverse  upper boundary $\bar{\rho}$ is of the order of characteristic size of a structure element of aerogel. The integral in the r.h.s. of Eq. (25) converges for large $k$, a principal contribution to this integral comes from the region of $k\sim (1/R)$. For $R\gg\xi_0$ this justifies approximation of $\eta_{jl}^{(s)}(\mathbf{k})$  by $\eta_{jl}^{(s)}(0)$.
The available experimental data for $S(k)$ are presented in arbitrary units, so that they can not be used directly for evaluation of $\bar{\tau}_{jl}^{(2)}$. Another possibility is to express $\bar{\tau}_{jl}^{(2)}$ in terms of $v(r)$, using the relation
$$
\int_0^{\infty} \bar{S}(k)dk=2\pi^2n\int_0^{\infty}v(r)rdr                     \eqno(26)
$$
and make a plausible guess about the form of the function $v(r)$, introducing in it adjustable parameters. Minimum two parameters are required. One is correlation radius $R$, introduced above, another is the overall amplitude $A$ so that $v(r)=A\bar{v}\left(\frac{r}{R}\right)$. In this case
$$
\bar{\tau}_{jl}^{(2)}=-\tau_{jl}^{(1)}\frac{5}{\zeta(3)}\frac{\pi^2}{9}I\cdot A\frac{R^2}{\xi_0 l_{tr}},                \eqno(27)
$$
where $I=\int_0^{\infty}\bar{v}(x)xdx$ is a number. This number  should not be very sensitive to the particular  form of $\bar{v}(x)$
A convenient model expression for $\bar{v}(x)$ in aerogel is \cite{frel,fomin}:
$$
\bar{v}(x)= \left[\frac{2}{\Gamma(D_f)}x^{D_f-3}-1\right]exp(-x),   \eqno(28)
$$
where $\Gamma(D_f)$ is Euler Gamma-function. This expression has a fractal asymptotic $\bar{v}(x)\sim x^{3-D}$ at $r\ll R$. At $r\gg R$  it tends to zero and it satisfies normalization condition $\int \bar{v}(x)x^2dx=0$.
For such $\bar{v}(x)$ $I=(3-D_f)/(D_f-1)$. If $D_f$=1.8 $I=3/2$. Experimental data and simulations \cite{parp2}  show that in aerogel impurities form clusters. This tendency corresponds to $D_f<$3 and positive $A$. In this case the sign of the correction $\bar{\tau}_{jl}^{(2)}$ is opposite to that of $\tau_{jl}^{(1)}$. It means that the destructive effect of impurities is weakened and the resulting $T_c$ is higher than that given by AG-theory. The increase of $T_c$ is due to adjustment of the condensate to local inhomogeneities which increases a gain of energy.

The correction $\bar{\tau}_{jl}^{(2)}$ is obtained with the aid of perturbation theory. It means that the inequality $|\bar{\tau}_{jl}^{(2)}|\ll\tau_{jl}^{(1)}$ has to be met. This condition sets an upper limit for $R$: $(AR^2/\xi_0 l_{tr})\ll 1$. A limit from below is set up by a condition securing dominant contribution of correlations: $nAR^2\bar{\rho}\gg 1$. So $R$ has to be within  a window $\frac{\bar{\rho}}{\xi_0}\ll \frac{AR^2}{\xi_0 l_{tr}}\ll 1$. The amplitude $A$ is not measured directly in experiments. For its estimation further model assumptions have to be used. Assuming, that threads of aerogel are "beads" formed by the balls of the radius $\bar{\rho}$ we arrive at a condition $(4\pi/3)An\bar{v}(3\bar{\rho})^3\approx\nu$, where $\nu$ is a "coordination number" i.e. average number of impurities touching a selected one. For "beads" $\nu\approx 2$.  If the fractal dimension $D_f\approx 2$ $A\sim\alpha(l_{tr}/R)$ with $\alpha\sim (1/10)$. Within this model we arrive at a stringent condition for applicability of a sum $\tau_{jl}^{(1)}+\bar{\tau}_{jl}^{(2)}$ as an approximation for $\tau_c$: $\bar{\rho}\ll\alpha R\ll\xi_0$. Particularly restrictive is the upper limit. Lifting of the formulated restriction requires an account of higher order terms in the series Eq. (8).
\section{Long-range correlations}
The series Eq. (8) is an expansion in the parameter $\frac{R^2}{\xi_0 l_{tr}}$.
 The $p$-th term of the series
$$
w^{(p+1)}_{jl}(\mathbf{k})=\eta_0^{p+1}n\int\frac{d^3k_1}{(2\pi)^3}...\frac{d^3k_p}{(2\pi)^3}G^{(0)}_{jm_1}(\mathbf{k_1})...
G^{(0)}_{m_pl}(\mathbf{k_p})
$$
$$
\left\langle
\sum_{a_1,...,a_p}\exp[i(\mathbf{k}-\mathbf{k_1})(\mathbf{r}_{a_1}-\mathbf{r}_e)]...\exp[i(\mathbf{k}_{p-1}-\mathbf{k}_p)
(\mathbf{r}_{a_p}-\mathbf{r}_e)]\right\rangle.          \eqno(29)
$$
Here $a_1,...,a_p,e$ are indices numbering impurities. Summation over the index $e$ has  rendered the factor $n$ in front of the integral. Expression in the angular  brackets is a product of values of the random function $\widetilde{S}(\mathbf{q})=\sum_{a}\exp[i\mathbf{q}(\mathbf{r}_{a}-\mathbf{r}_e)]$  taken at different values of its argument. Essential contribution to the integral comes from the region $k_s\sim(1/R)$. If $R$ is much greater than the average distance between the impurities the number of impurities contributing coherently to the sum is large. In this case the random function $S(\mathbf{q})$ is close to its average value i.e. to the structure factor $\widetilde{S}(\mathbf{q})=\langle S(\mathbf{q})\rangle$ and in the leading order over $nR^3$ the result of averaging in Eq. (29) can be represented as a product
$S(\mathbf{k}-\mathbf{k_1})S(\mathbf{k_1}-\mathbf{k_2})...S(\mathbf{k}_{p-1}-\mathbf{k}_p)$. Then the consecutive members of the series for $W(\mathbf{k})$ are related via
$$
w^{(p+1)}_{jl}(\mathbf{k})=\eta_0\int\frac{d^3k_1}{(2\pi)^3}S(\mathbf{k}-\mathbf{k_1})G_{jm}(\mathbf{k_1})w^{(p)}_{ml}(\mathbf{k_1}) \eqno(30)
$$
and summation of the series renders the integral equation for $W_{jl}(\mathbf{k})$:
$$
W_{jl}(\mathbf{k})=n\eta_0\delta_{jl}+\eta_0\int\frac{d^3k_1}{(2\pi)^3}S(\mathbf{k}-\mathbf{k_1})G_{jm}(\mathbf{k_1})W_{ml}(\mathbf{k_1}).  \eqno(31)
$$
for substitution in Eq. (10) we need only $W_{jl}(\mathbf{k}=0)$. The main contribution to the integral in the r.h.s. of Eq (31) comes from the region $k\leq (1/R)$. At $W=const.$ the integral converges, so that $W_{jl}(\mathbf{k})\approx W(0)\delta_{jl}$ can be used as an approximate solution of Eq. (31). For an isotropic $W_{jl}(\mathbf{k})$ the Green unction $G_{jm}(\mathbf{k_1})$ can be averaged over directions of $\mathbf{k}_1$, rendering $\bar{G}_{jm}(\mathbf{k_1})=-\frac{7}{9\xi_s^2k^2_1}\delta_{jm}$
Then the value of $W(0)$ is determined by Eq. (31):
$$
W(0)=\frac{n\eta_0}{1-\eta_0 Q},              \eqno(32)
$$
where $Q=\int\frac{d^3k_1}{(2\pi)^3}S(\mathbf{k_1})\bar{G}(\mathbf{k_1})$. Comparison with Eq.(18) shows, that summation of the perturbation series reduces to substitution instead of $\sigma_{tr}^{(i)}$ an ``effective'' cross-section, which takes into account correlation of positions of scattering centers.
For the model correlation function Eq. (28) with an account of Eqns. (10),(16)
$$
\tau_c=\frac{\pi^2}{4}\frac{\xi_0}{l_{tr}}\frac{1}{1+\frac{\pi^2}{4}B(D)\frac{R^2}{\xi_0l_{tr}}A},           \eqno(33)
$$
where $B(D)=\frac{10\Gamma(D)}{9\zeta(3)}\frac{3-D}{D-1}$.

The obtained expression (33) for a relative lowering of $T_c$ has a similar structure to that, suggested on a basis of heuristic argument by Sauls and  Sharma \cite{sauls2}. In a limit $\frac{R^2}{\xi_0l_{tr}}A\ll 1$ $\tau_c$ is proportional to the "pairbreaking parameter" $\frac{\xi_0}{l_{tr}}$, as in the conventional theory. In the opposite limit the mean free path cancels out and  $\tau_c\approx\frac{\xi_0^2}{R^2 AB(D)}$ is determined by  geometric characteristics of aerogel. When $R$ is growing $\tau_c$ decreases and $T_c$ tends to its bulk value. If $A$ is estimated within the "model of beads" the dependence on $R$ is different: $\tau_c\approx\frac{\xi_0^2}{l_{tr}R}\frac{1}{\alpha B(2)}$
In $^3$He $T_c$ and consequently $\xi_0$  strongly depend on pressure. At solidification pressure $\xi_0$ is approximately four times smaller than at the vapor pressure. In 98\% silica aerogel both limits of Eq. (33) can be reached. With the aid of their interpolation Sauls and  Sharma \cite{sauls2} were able to fit the pressure dependence of $T_c$ in such aerogel for the interval of pressures from 6 to 32 bars, using realistic parameters of aerogel.  Within the region of validity of Ginzburg and Landau equations Eq. (33) renders a theoretical justification to their interpolation.

$^3$He in aerogel was used here as a simple example, where effect of correlations can be analyzed and compared with the existing data. To obtain a concrete result in deriving Eq. (33) specific properties of correlations in aerogel are used, so it can not be directly applied to a different object. On the other hand the argument of previous section, leading to Eq. (31) is general and can be used at analyzing of effect of correlated impurities in metallic superconductors.

\section{Discussion}
The example of superfluid $^3$He in aerogel clearly demonstrates that correlations of impurities can substantially weaken their pairbreaking effect and increase the transition temperature with respect to that expected for non-correlated impurities. More literally, it shows that the description of aerogel as a uniform continuous media is not sufficient.  Correlations introduce a spatial scale $R$. Adjustment of a superfluid condensate to the external inhomogeneity increases its gain of energy in comparison with a uniform condensate.  Mostly important are correlations with a characteristic radius $R$ of the order or exceeding the superfluid coherence length $\xi_0$. Superfluid condensate is more susceptible to such correlations, because reaction of condensate on external perturbation is determined by momenta on the order of  $1/\xi_0$. For comparison, reaction of normal quasi-particles is determined by much higher momenta - on the order of Fermi momentum. In this case effect of correlations renders only a small correction to the mean free path. Enhanced susceptibility of condensate to long range correlations manifests itself below  $T_c$ as well. A separate analysis shows that in a superfluid phase correlations effect a temperature dependence of the order parameter because of the interplay of correlation radius and Ginzburg and Landau correlation length $\xi(T)$ \cite{fom3,fom-sur}.

In the last few years experimental activity in $^3$He shifted to using different types of aerogel. On one hand there are highly anisotropic "nematic" aerogels \cite{askh,dm4} on the other -- highly isotropic silica aerogels, exhibiting new features \cite{halp2}. Application of the present approach to these objects can bring further interesting results. Practically even more useful and stimulating would be a systematic experimental investigation of the effect of correlated impurities in metallic unconventional superconductors.

\section{Acknowledgments}
It is my great pleasure and honor to make contribution to the issue, celebrating 150 years of Margarita and Vladimir Man`ko.
It is also good opportunity to congratulate my friends with this impressive date and to wish them many more happy years in science and in ordinary life.

This work was supported in part by the Russian Foundation for Basic Research, project \# 14-02-00054-a

\end{document}